\documentclass[aps,prb,twocolumn,groupedaddress,floatfix]{revtex4-2}

\usepackage{graphicx}
\usepackage{dcolumn}
\usepackage{bm}
\usepackage{mathtools}
\usepackage{xcolor}

\usepackage{physics}

\newlength\mylen  \setcitestyle{citesep={,\kern-\mylen}}

\begin{document}

\title{New Source of Spin-hot spot in displaced silicon double quantum dots  }
\author{
Sanjay Prabhakar
}
\affiliation{
$^1$Department of Natural Sciences, Dean L. Hubbard Center for Innovation, Northwest Missouri State University, 800 University Drive, Maryville, MO 64468\\
$^2$MS2Discovery Interdisciplinary Research Institute, Wilfrid Laurier University, Waterloo, ON, N2L 3C5 Canada
}

\affiliation{
$^{*)}$Author to whom correspondence should be addressed: sanjay@nwmissouri.edu}

\date{Nov 6, 2023}

\begin{abstract}
Controlling electron spins in double quantum dots allows individual electrons to be trapped and manipulated  for next-generation solid-state qubit devices.
In this paper, the study analyzes spin relaxation due to deformation potentials of acoustic phonon in single and double quantum dots under in-plane and out-of-plane magnetic fields, showing that in single quantum dots the relaxation rate is highly sensitive to low in-plane magnetic fields ($<1T$) but converges near a spin-hot-spot region. In a single quantum dot, the spin-hot spot arises from well-understood level crossings between singlet and triplet states. In double quantum dots, a new and unusual spin-hot spot appears as the dots are pulled apart from the origin, with spin-relaxation rates three orders of magnitude lower than conventional single quantum dots. In displaced quantum dots dominated by magnetic confinement, two distinct spin-hot spots appear at different in-plane magnetic field strengths, where spin-relaxation time varies from millisecond to picosecond. When quantum dots are separated by about 60 nm, calculations predict oscillations in spin-hot spots as the in-plane magnetic field changes. These unusual spin-hot spot oscillations occur at low magnetic fields ($<1T$), resulting in spin-relaxation rates about four orders of magnitude lower than those of conventional high-field spin-hot spots ($\approx 4.5T$). The extremely low spin-relaxation rate at the spin-hot spot enables the preparation of qubit superposition states for quantum computing and information processing.
\end{abstract}

\maketitle
\section{Introduction}
Gate-controlled manipulation of single and double electron spins in quantum dots using electric and magnetic fields is a key ongoing effort toward solid-state quantum computing and quantum information processing~\cite{sobrino24,khomitsky25,bhat22,suarez25,banerjee25,raith12,raith11,bulaev07,sp24,zwerver23,paquelet23}. Single and double quantum dots, created using advanced techniques like molecular beam epitaxy and metal oxide chemical vapor deposition  with lithographic processes in metal oxide field effect transistors, are crucial for qubit gate operations, enabling initialization of one or two qubits from source to drain \cite{wang22,fernandez22,zhou23,jock18,burkard23,steinacker25,woods23}. Advanced technology enables rapid spin-current measurements in silicon quantum dots, which are preferred over GaAs due to having small spin-orbit coupling effects, resulting in longer qubit coherence times for more effective gate operations \cite{gilbert23,liles24,jock18}. In the lab, long spin-relaxation ($T_1$) and decoherence ($T_2\approx 2T_1$) times can be achieved through mechanisms like electron-phonon interaction, electron-hyperfine interactions in presence of spin-orbit coupling, allowing qubit spins to be reliably initialized, manipulated, and read out which is key requirement for quantum computing \cite{wang24,appel22,haldar25,kedim25,dyte25,kazemi26,zhang26,memon24,oakes23}.

In silicon quantum dots, weak spin-orbit coupling is incorporated into the Hamiltonian via Rashba and linear Dresselhaus terms\cite{raith12,jock18,debnath24,ferreira23,hetenyi22}. The Rashba spin-orbit coupling originates from structural inversion asymmetry along the growth direction, while the Dresselhaus spin-orbit coupling arises from bulk inversion asymmetry of the crystal lattice. Spin relaxation from electron–phonon interactions (piezo and acoustic) in quantum dots creates spin-hot spots through level crossings between spin and orbital states\cite{raith12,sp24,bulaev07,wang24,hosseinkhani22}. In silicon quantum dots, only acoustic phonons significantly contribute to spin flips, unlike in III-V semiconductors where piezo-phonons are strong\cite{kolok25,coste23,han22,sp13,sp24}. Spin relaxation in silicon quantum dots has been extensively studied\cite{hsueh23,bosco22,hosseinkhani22,liu22,blumoff22}. Spin-relaxation in double silicon quantum dots has recently been observed experimentally for quantum computing applications\cite{lundberg20,ciriano21,crepieux22,ginzel23}. Experiments show that in silicon double quantum dots, two qubit spins can be initialized in singlet states and driven into superposition at spin-hot spots using rapid adiabatic magnetic field pulses\cite{zhao22,krzywda20,takeda24}. The  present paper reports computational results on acoustic phonon induced spin relaxation and identifies multiple spin-hot spots suitable for generating two-qubit superposition states applications in quantum information processing.

\begin{figure}
\includegraphics[width=8.5cm,height=6cm]{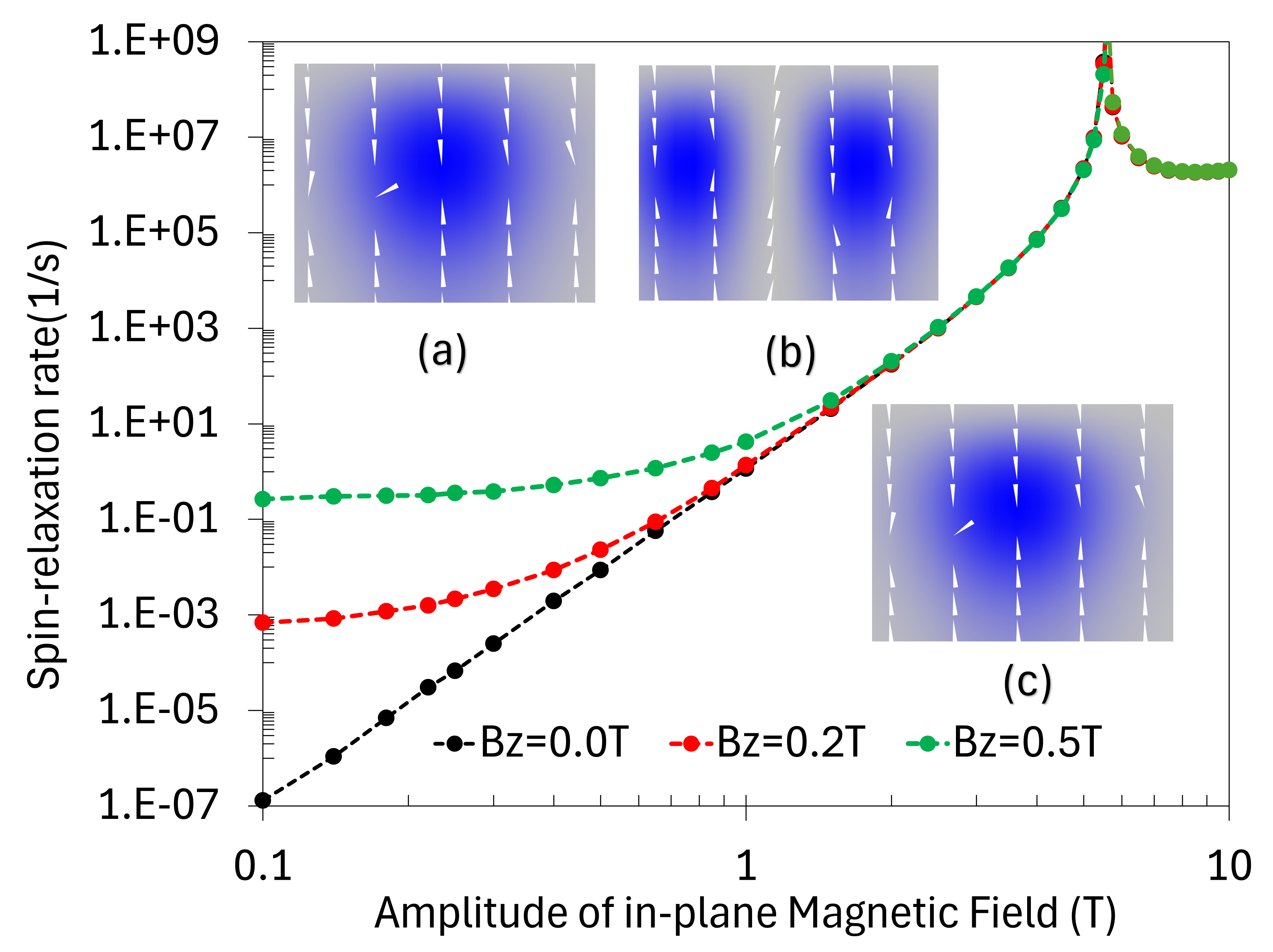}
\caption{\label{fig1} (Color online) Spin-relaxation rate versus amplitude of in-plane magnetic field in a single quantum dot in presence of out-of-plane magnetic fields at $B_0=0.0T$ (black), $0.2T$ (red), and $0.3T$ (green). Spin-hot spot can be observed in the vicinity of $5.5T$ of amplitude of in-plane magnetic fields, where spin relaxations are immune to out-of-plane magnetic fields, $B_0> 1.5$. When $B_0<1.5T$, spin-relaxations are very sensitive to the out-of-plane magnetic fields. Insets (a), (b) and (c) are the probability density of three lowest electron spin states at the spin-hot spot ($B_{in}=5.75T, B_0=0.2T$) along with kinetic momentum density (white surface arrow). We chose $\ell_0=25nm$.}
\end{figure}

\begin{figure*}
\includegraphics[width=18cm,height=7cm]{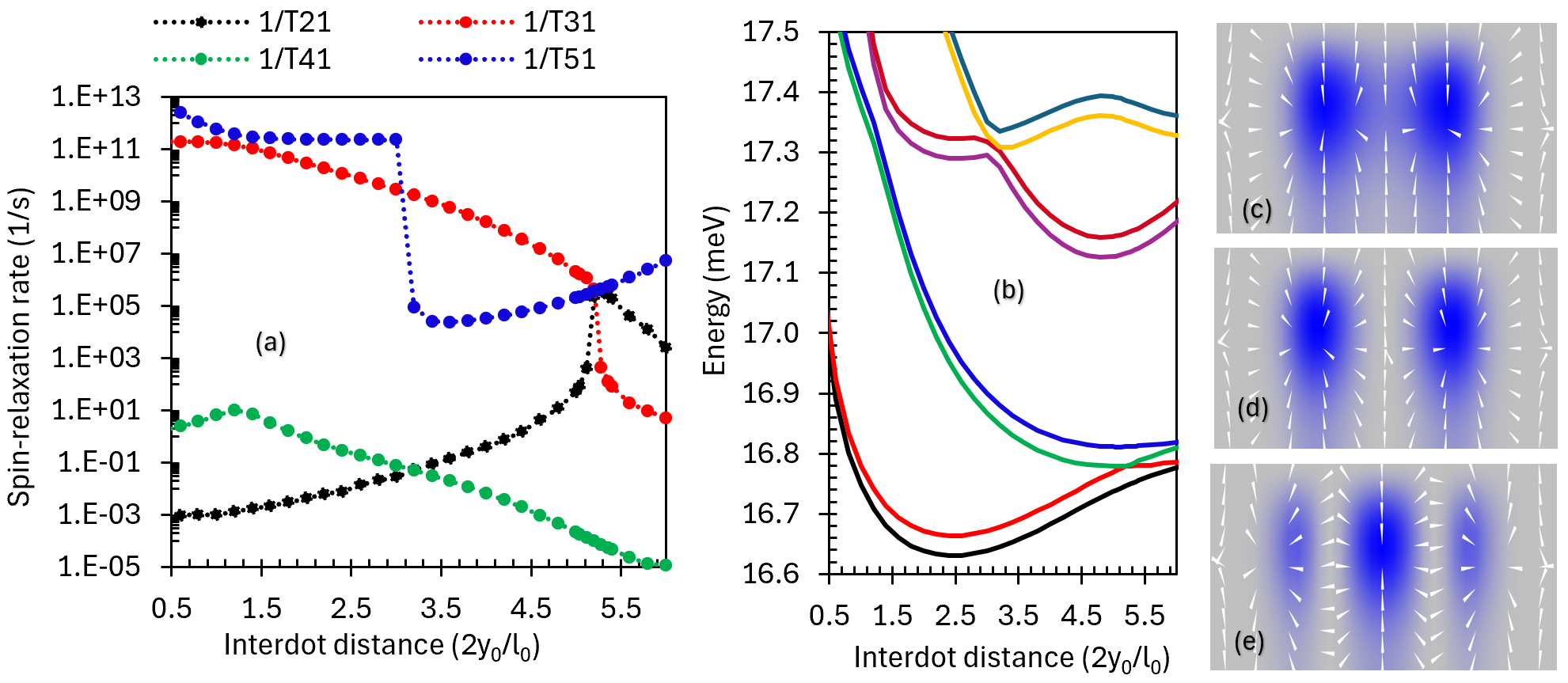}
\caption{\label{fig2} (Color online) Spin-relaxation rate versus inter-dot distances between double quantum dot is shown in (a). Spin-hot spot, so far not observed in the literature, can be seen in the vicinity of quantum dot displaced to $32nm$ from the origin due to level crossing of the bands seen in (b). The probability densities of ground, 3rd and 5th electron spin states at the spin-hot spot along with kinetic momentum density (white surface arrow) are shown in (c,d,e). The probability density of second and fourth states of double quantum dots looks exactly the same to (a) and (b) with slight variations due to spin-splitting behavior in presence of magnetic field, as seen in the inset plot of Fig.~\ref{fig1} (a) and (c).  We chose $\ell_0=25nm$, $B_0=0.2T$. In (b) and (c), we also chose $B_{in}=0.2T$ and $y_0=60nm$.}
\end{figure*}

Spin-hot spots arise from strong mixing of singlet and triplet spin and orbital states, leading to enhanced phonon-mediated spin relaxation in admixed qubit spin states, which has been widely studied~\cite{jock22,spence22,sp13,sp24,raith12,bulaev07}. In Ref.~\cite{raith11}, spin-hot spots are observed in double quantum dots when an out-of-plane magnetic field is applied, while in Ref.~\cite{bulaev07}, two spin-hot spots are observed when in-plane and out-of-plane magnetic fields are applied. The present paper identifies a new source of spin–hot-spot mechanism that emerges when in-plane and out-of-plane magnetic fields are applied as the two quantum dots are pulled apart from their origin, with spin-relaxation rates about four orders of magnitude smaller than those driven by magnetic-field variation alone (Figs. \ref{fig1},\ref{fig2}). In a displaced quantum dot where magnetic confinement potentials dominates over gate potential, two spin-hot spots appear at different in-plane magnetic field strengths. In displaced quantum dots with about $60$nm separation, spin-hot spots oscillate and a third spin-hot spot appears due to stronger magnetic confinement from combined in-plane and out-of-plane magnetic fields (Fig.\ref{fig4}). In Ref.\cite{jock22,jock18}, the authors propose preparing two qubits in a singlet state and reading out superposition states, exploiting singlet–triplet overlap induced by rapid adiabatic tuning of the detuning parameter (magnetic field). Hence, spin-hot spots provide ideal qubit readout states, where preserving superposition information critically depends on decoherence times with $T_2\approx T_1$. This paper demonstrates the existence of multiple spin-hot spots, some with decoherence times up to four orders of magnitude longer than others, making them ideal for qubit manipulation in quantum information processing.

The paper is organized as follows. Section \ref{theoretical-model} presents a theoretical model describing spin relaxation in single and double silicon quantum dots arising from interactions with acoustic phonons. Section \ref{computational-method} briefly outlines a computational diagonalization approach using finite element method simulations to determine the energy spectrum and evaluate the matrix elements governing phonon-mediated spin relaxation rates.
Section \ref{results-and-discussions} presents and analyzes results on electron spin relaxation in single and double silicon quantum dots, driven by spin–orbit coupling with acoustic phonons under both in-plane and out-of-plane magnetic fields. Finally, Section \ref{conclusion} provides a concise summary of the main results of the study.

\begin{figure*}
\includegraphics[width=18.0cm,height=7cm]{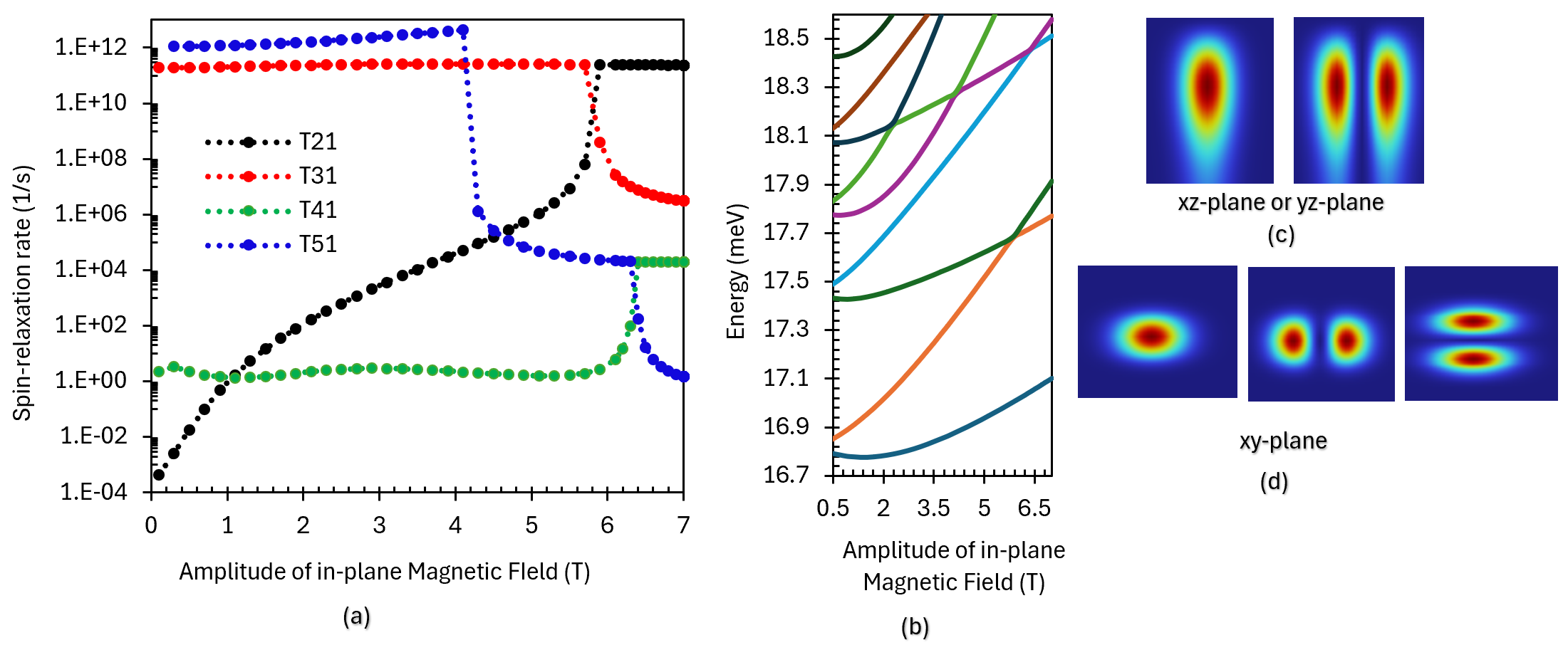}
\caption{\label{fig3} (Color online) (a) Spin-relaxation rate versus amplitude of in-plane magentic field in a double quantum dots separated at $y_0=10nm$ from the origin. Very unusual two spin-hot spot can be observed at $5.9T$ and $6.4T$ due to level crossing of the bands that is shown in Fig.(b).  Probability densities of ground state, and 2nd excited state of the double dots in the xz-plane or yz-plane are shown in Fig.(c). Probability densities of ground state, 2nd excited state and 4th excited state of the double dots in the xy-plane  are shown in Fig.(d). We chose $\ell_0=25nm$, $B_0=0.2T$. In (c) and (d), we also chose $B_{in}=0.2T$}.
\end{figure*}

\begin{figure*}
\includegraphics[width=15.0cm,height=9cm]{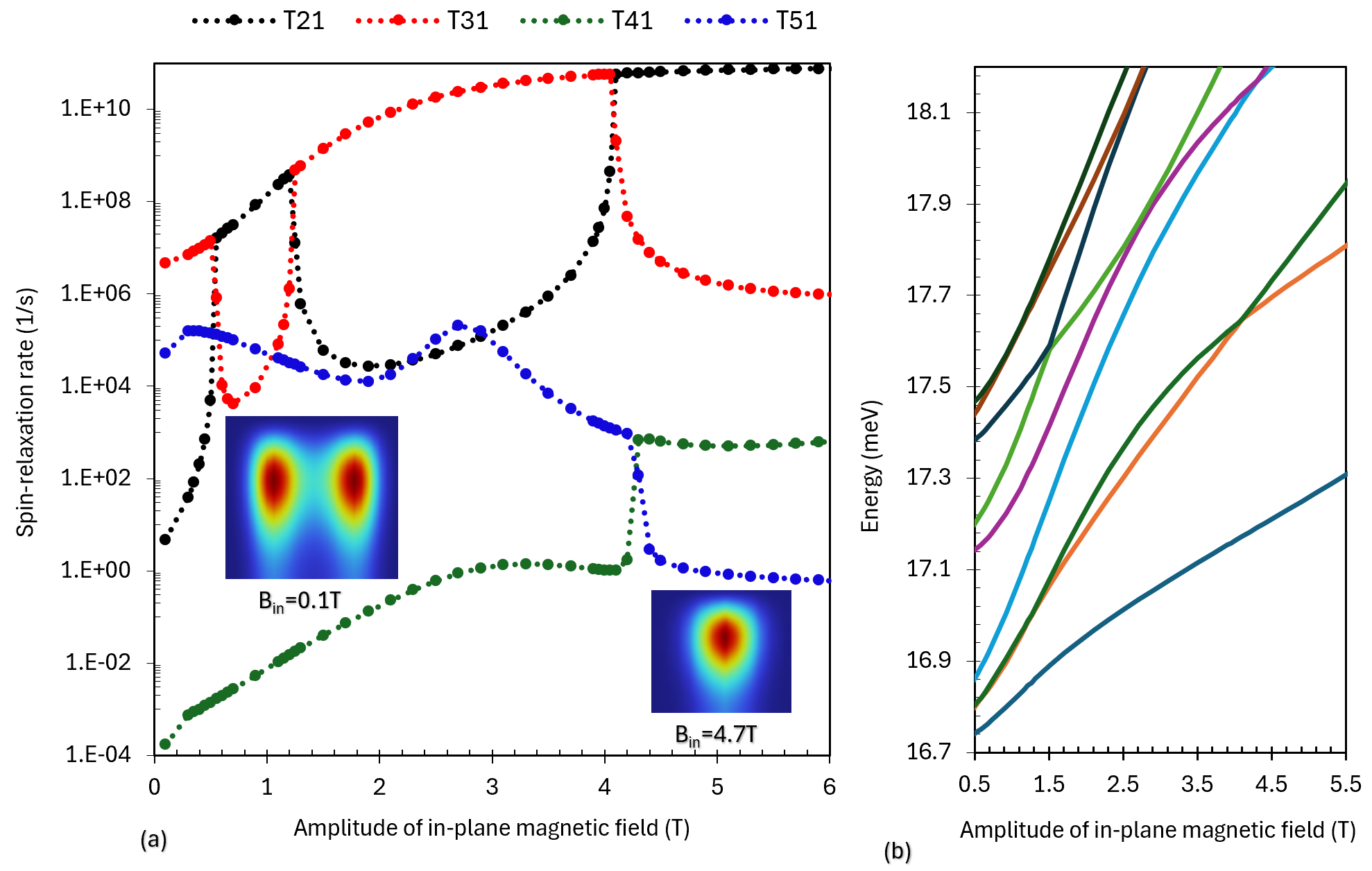}
\caption{\label{fig4} (Color online) (a) Spin-relaxation rate versus amplitude of in-plane magnetic field in a double quantum dots separated at $y_0=60nm$ from the origin. In addition to the unusual two spin-hot spot at $5.9T$ and $6.4T$, oscillating spin-hot spot cat $B=0.5$ and $1.2T$ can be observed due to level crossing of the bands that is shown in Fig.(b).  Probability densities of ground state, and 2nd excited state of the double dots in the xz-plane or yz-plane are shown in Fig.(c). Probability densities of ground state, 2nd excited state and 4th excited state of the double dots in the xy-plane  are shown in Fig.(d). We chose $\ell_0=25nm$, $B_0=0.2T$. In (c) and (d), we also chose $B_{in}=0.2T$}.
\end{figure*}

\section{Theoretical Medel}\label{theoretical-model}
The study models an electron’s three-dimensional motion in a quantum dot under both in-plane and out-of-plane magnetic fields, incorporating the Rashba and linear Dresselhaus spin–orbit couplings. Following established approaches from prior studies \cite{corrigan23,raith11,raith12,bulaev07,sp13,sp24}, the total Hamiltonian can be written as
\begin{equation}
H = H_0 + H_1 + H_R + H_D, \label{H}
\end{equation}
\begin{eqnarray}
H_0 &=& \frac{P_x^2}{2m_x} + \frac{P_y^2}{2m_y}+\frac{P_z^2}{2m_z} + \frac{1}{2}m \omega_0^2 V(x,y) +eEz, ~~~ \label{H0} \\
 H_1 &=& \frac{1}{2}\mu_B \left(g_x B_x \sigma_x + g_y B_y \sigma_y + g_z B_z \sigma_z \right), ~~~ \label{Hz} \\
 H_R &=& \alpha_R\left(\sigma_xP_y - \sigma_y P_x\right),\label{HR}\\
  H_D &=& \alpha_D\left(-\sigma_xP_x + \sigma_yP_y\right).\label{HD}
\end{eqnarray}
Here, $\mathbf{P}=\mathbf{p} + e \mathbf{A}$ with the vector potential, $\mathbf{A}= \left(-y B_z/2, xB_z/2, yB_x-xB_y\right)$ and $\mathbf{p}$ is the momentum operator. $V_z=Ez$ is the confinement potential in z-direction due to band offset of semiconductor material at the interface of the heterojunction and $V(x,y)$ is the confinement potential in the lateral direction.   $H_1$ is the Zeeman spin-splitting energy due to  applied in-plane magnetic fields, $Bx$ and $B_y$, and out-of-plane magnetic field, $B_z$. Here, $m_i (i=x,y,z)$ is the effective mass of electron in the QDs, $\ell_0 = \sqrt{\hbar/m\omega_0}$ is the radius of the lateral size of the dots and $g_i (i=x,y,z)$ is the bulk g-factor of electron. The spin-orbit coupling, $H_{so} = H_R+H_D$, where $H_R$ is the Rashba spin-orbit coupling  and $H_D$ is the linear Dresselhaus spin-orbit coupling.
Also, $\sigma_x$, $\sigma_y$ and $\sigma_z$ are the Pauli spin matrices. The lateral confinement potetnial, $V_{xy}$, of the single and double quantum dot can be written as
\begin{eqnarray}
V(x,y) &=& x^2+y^2, single~ QDs \label{vxy1} \\
V(x,y) &=& x^2+\frac{1}{4y_0^2} (y^2-y_0^2)^2, double~ QDs. \label{vxy1}
\end{eqnarray}
The authors of this paper shift their study in evaluating the phonon-induced spin relaxation rate in quantum dots at low temperatures, where acoustic phonons dominate and thermal excitations are minimal. The calculation aims to provide theoretical descriptions of how electron–phonon interacts, together with spin–orbit coupling, lead to spin-flip transitions between quantum dot energy levels, thereby determining the spin relaxation time relevant for qubit decoherence. Following Refs.~\cite{raith11,raith12,prada08}, the interaction between electron and acoustic phonon can be written as
\begin{equation}
u^{\mathbf{q}\lambda}_{ph}\left(\mathbf{r}\right)=i \sum_{\lambda=l,t} \sqrt{\frac{\hbar}{2\rho V \omega_{\mathbf{q}\alpha}}} e^{i \mathbf{q\cdot r} } q D_q^\lambda b^{\dag}_{\mathbf{q}\alpha} + H.c.,
\label{u}
\end{equation}
\begin{equation}
D_q^\lambda= \Xi_d \mathbf{\hat{e}_q^\lambda \cdot \hat{q}} + \Xi_u \hat{e}_{q_z}^\lambda  \hat{q_z},
\label{Dq}
\end{equation}
where $\rho$ is the crystal mass density, $V$ is the volume of the dot.  Also, $b^{\dag}_{\mathbf{q}\alpha}$ creates an acoustic phonon with wave vector $\mathbf{q}$ and polarization $\mathbf{\hat{e}_q^\lambda}$, where $\lambda=l,t_1,t_2$ are chosen as one longitudinal and two transverse phonon modes. The polarization directions of the induced phonon are $\mathbf{\hat{e}_q}^l=\left(\sin\theta \cos\phi, \sin\theta \sin\phi, \cos\theta \right)$,   $\mathbf{\hat{e}_q}^{t_1}=\left(-\sin\phi, \cos\phi, 0 \right)$, and $\mathbf{\hat{e}_q}^{t_2}=\left(\cos\theta \cos\phi, \cos\theta \sin\phi, -\sin\theta \right)$. The deformation potential strength is determined by the dilatation potential constant, $\Xi_d$ and shear potential constant, $\Xi_u$.  Based on the Fermi Golden Rule, the phonon induced spin transition rate is given by~\cite{raith11,raith12,prada08}
\begin{equation}
w_0=\frac{2\pi}{\hbar}\int \frac{d^3\mathbf{q}}{\left(2\pi\right)^3}\sum_{\alpha=l,t}\arrowvert M\left(\mathbf{q}\alpha\right)\arrowvert^2\delta\left(\hbar s_\alpha \mathbf{q}-\varepsilon_{f}+\varepsilon_{i}\right),
\label{1-T1}
\end{equation}
where  $s_l$,$s_t$ are the longitudinal and transverse acoustic phonon velocities in Si QDs.  The matrix element $M\left(\mathbf{q}\lambda\right)=\langle \psi_i|u^{\mathbf{q}\lambda}_{ph}\left(\mathbf{r},t\right)|\psi_f\rangle$ with the emission of one phonon has been calculated numerically \cite{comsol}. Here $|\psi_i\rangle$ and $|\psi_f\rangle$ correspond to the initial and finial states of the Hamiltonian $H$. More precisely, we write the spin-relaxation rate due to longitudinal and transverse phonon as
\begin{widetext}
\begin{eqnarray}
w_{0l}&=& \frac{\Delta E^5} {210\pi\rho\hbar^6 s_l^7}
\begin{bmatrix}
\left(35 \Xi_d^2 + 3 \Xi_u^2 + 14 \Xi_u \Xi_d \right) \left(|M_x|^2 + |M_y|^2\right)
 + \left(35 \Xi_d^2 + 15 \Xi_u^2 + 42 \Xi_u \Xi_d \right)|M_z|^2
\end{bmatrix},
\label{W01}\\
~\nonumber\\
w_{0t}&=& \frac{\Delta E^5 \Xi_u^2} {105\pi\rho\hbar^6 s_t^7}
\begin{bmatrix}
2\left(|M_x|^2 + |M_y|^2\right) + 3 |M_z|^2
\end{bmatrix}.
\label{W0t}
\end{eqnarray}
\end{widetext}
where $M_x = \langle \psi_i|x|\psi_f\rangle$, $M_y=\langle \psi_i|y|\psi_f\rangle$, and $M_z=\langle \psi_i|z|\psi_f\rangle$.

\section{Computational Method}\label{computational-method}

We suppose that a  QD is formed at the center of a $1200\times 1200~\mathrm{nm^2}$ geometry.  We then diagonalize the total Hamiltonian $H$ numerically using the Finite Element Method~\cite{comsol}.  Since the geometry is much larger compared to the actual lateral size of the QD, we impose Dirichlet boundary conditions, find the  eigenvalues, eigenfunctions and the matrix elements $M\left(\mathbf{q}\alpha\right)$ of $H$. The results presented in Fig.\ref{fig1} is consistent with the Fig.5 of Ref.\cite{raith11} for single quantum dots. The material constants of silicon for the simulations are taken as follows.  We use the parameters of a silicon quantum dots grown along the [001] direction with silicon dioxide on the top. The two-dimensional electron gas is defined at the heterojunction of $Si-SiO_2$ layer. The effective masses are, $m_{xy}=0.19$ and $m_z=0.98$ for silicon and $m_{ox}=0.42$ for silicon-oxide layer. The bulk g-factor is chosen to be $g_0=2$. The electric field strengths that control the Rashba spin-orbit coupling is $3.1MV/m$ and the Dresselhaus spin-orbit coupling is $1.0MV/m$. The choices of these electric field strengths make the Dresselhaus spin-orbit coupling strength about three times larger than the Rashba spin-orbit coupling, which is consistent to the experimental results in Ref.~\cite{jock18}. To calculate the spin-relaxation spin due to acoustic phonon, we use $ s_l = 9150 m/s$ for LA phonons, and  $s_tt = 5000 m/s$ for TA phonons, $\rho = 2330 kg/m^3$. The choice of deformation potential constants is consistent to Ref.\cite{raith11} as $Xi_d = 5eV$ and $Xi_u =9eV$.

\section{Results and Discussions}\label{results-and-discussions}

In Fig.\ref{fig1}, we plot the spin-relaxation rate vs amplitude of in-plane magnetic field  in presence of  out-of-plane magnetic fields, $B_z=0.0T, 0.2T, 0.5T$ in a single quantum dot.  Spin-hot spot can be observed in
the vicinity of $5.5T$ of amplitude of in-plane magnetic fields, where spin relaxations are immune to out-of-plane magnetic fields, $B0 > 1.5$. When $B0 < 1.5T$, spin-relaxations are very sensitive to the out-of-plane magnetic fields. In the Inset plots of Fig.\ref{fig1} (a), (b) and (c), we plot the probability density of three lowest electron spin states at the spin-hot spot ($Bin = 5.75T, B0 = 0.2T$) along with kinetic momentum density (white surface arrow). Clearly we observe level crossing between spin and orbital states in Fig\ref{fig1}(b). However, there is no any clear pattern on spin-relaxation rate vs in-plane magnetic field in presence of out-of-plane magnetic fields in z-direction, indicating cumbersome mathematical calculations of phonon interaction with quantum dots.

In Fig.\ref{fig2}, we plot various spin-relaxation rate vs inter-dot distances  in presence of  out-of-plane and in-plane magnetic fields, $B_z=0.2T$ and $B_{in} = 0.2T$, respectively in a double quantum dot. Unusual spin-hot spot can be observed in the vicinity of $2y_0/\ell_0=4.8$ or $y_0=32nm$. Note that $y_0$ is a parameter that controls the separation of the dots along y-direction from dots' center. The data shown in green suggests that spin-relaxation rate decreases, which enhances the decoherence time for qubit gate operation, when opposite spin-states between left and right dot make transition. In Figs.2(b,c,d), we plot the probability density of first three spin states along with kinetic momentum operator when center of dots in y-direction are separated by $60nm$. In the probability of ground state in (b), we clearly observe two  dots separation with some of the wavefunctions overlap. In (c), we clrealy observe different probability density than (a) due to the level crossing of the bands among the dots. The crossing of the band structures due to the separation of the dots are shown in Fig. (e).

In Ref.\cite{raith11}, authors investigate the bandcrossing as double quantum dots are pulled apart from the origin in presence of zero magnetic field. In Fig.\ref{fig2}, we find that splitting of bands are possible and hence able to find the unusual spin-hot spot when dots are spearated at about 60nm in presence of both in-plane and out-of-plane mannetic fields. In Ref.~\cite{bulaev07}, authors investigate the influence of in-plane and out-of-plane mangetic fields in single heavy hole quantum dots, though it is not quite clrer how in-plane magnetic fields that has z-component in the vector potential is implemented in the two-dimensional Hamiltonian. Hence we consider complete Hamiltonian \ref{H} and consider canonical momentum, $p_z=\hbar k =0$ at $\Gamma$-point and write the two dimensional Hamiltonian,
\begin{equation}
H = H_{0} + H_2  +  H_R+H_D,
\label{H-1}
\end{equation}
where in-plane magnetic field components ($B_x, B_y$) can be incorporated in $H_2$ and  $g_z > g_{x,y}$ \cite{raith11}. The terms $H_R$ and $H_D$ are same as in Eq.\ref{H} and remaining terms are
\begin{equation}
H_{0} = \frac{P_x^2+P_y^2}{2m}  + {\frac{1}{2}} m \omega_o^2 min\left\{ \left(r-r_0\right)^2, \left(r+r_0\right)^2 \right\}+ \frac{\Delta}{2}  \sigma_z,
\label{hxy-1}
\end{equation}
\begin{equation}
H_{2} = \frac{e^2}{2m_z}\left( B_y x - B_x y \right)^2,
\label{H2}
\end{equation}
where $\Delta=g_z\mu_B B_z$. At this point,  we introduce the relative coordinate $\mathbf{R}=\mathbf{r}-\mathbf{r_0}=\left(X,Y,0\right)$ and  the relative momentum $\mathbf{P}=\mathbf{p}-\mathbf{p_0}=\left(P_X,P_Y,0\right)$, where $\mathbf{p_0}=m\mathbf{\dot{r}}_0$ is the momentum of the displaced dot which vanishes due to the fact that the displaced double dots from the origin are stationary.  The displaced dots of Hamiltonian~(\ref{H-1}) in terms of relative-coordinate and momentum can be written as
\begin{equation}
H=H_0\left(\mathbf{P},\mathbf{R}\right)+H_{so}\left(\mathbf{P},\mathbf{R}\right)+
H_{ad}\left(\mathbf{P},\mathbf{R};\mathbf{p_0},\mathbf{r_0}\right),
\label{total-adiabatic}
\end{equation}
where
\begin{eqnarray}
H_0\left(\mathbf{P},\mathbf{R}\right) = {\frac {1}{2m}}\left\{\mathbf{P}+e\mathbf{A}(\mathbf{R})\right\}^2 + {\frac{1}{2}} m \omega_o^2R^2 + {\frac \Delta 2}  \sigma_z,\label{tilde-hxy}~~~
\end{eqnarray}
\begin{eqnarray}
 H_R\left(\mathbf{P},\mathbf{R}\right) &=& \alpha_R\left(\sigma_x \left\{P_Y+eA_Y\right\} - \sigma_y \left\{P_X+e A_X\right\}\right),\label{HR-1}~~~\\
  H_D\left(\mathbf{P},\mathbf{R}\right) &=& \alpha_D\left(-\sigma_x  \left\{P_X+e A_X\right\} + \sigma_y \left\{P_Y+eA_Y\right\} \right).~~~\label{HD-1}
\end{eqnarray}
\begin{eqnarray}
H_{ad}\left(\mathbf{P},\mathbf{R};\mathbf{p}_0,\mathbf{r}_0\right)&=&\frac{1}{m}\left\{\mathbf{P}
+e\mathbf{A}\left(\mathbf{R}\right)\right\}\cdot \left\{\mathbf{P}+e\mathbf{A}\left(\mathbf{R}\right)\right\}\nonumber\\
&&+H_{so}\left(\mathbf{p}_0,\mathbf{r}_0\right),\label{Had}
\end{eqnarray}
The  Hamiltonian $H_0\left(\mathbf{P},\mathbf{R}\right)$, $H_R\left(\mathbf{P},\mathbf{R}\right)$, $H_D\left(\mathbf{P},\mathbf{R}\right)$ can be diagonalized on the basis of the number states $|n_+,n_-,\pm 1\rangle$, which is the same for left and right dots:
\begin{equation}
H_0\left(P,R\right)=\left(N_+ + \frac{1}{2}\right)\hbar\Omega_+ + \left(N_- + \frac{1}{2}\right)\hbar\Omega_-  + \frac{\Delta}{2}\sigma_z,\label{H0}
\end{equation}
where $N_{\pm}=a^\dagger_{\pm}a_{\pm}$ are the number operators with eigenvalues $n_\pm \in N_0$. Here,
\begin{eqnarray}
a_{\pm}=\frac{1}{\sqrt{4m\hbar\Omega}}\left(iP_x \pm P_y\right)+\sqrt{\frac{m\Omega}{4\hbar}}\left( X \mp i Y \right),\\ \label{a-lowering}
a_{\pm}^\dagger=\frac{1}{\sqrt{4m\hbar\Omega}}\left(-iP_x \pm P_y\right)+\sqrt{\frac{m\Omega}{4\hbar}}\left( X \pm i Y \right),\label{a-raising}
\end{eqnarray}
provided that $\left[a_{\pm},a_{\pm}^\dagger\right]=1$.
Correspondingly, the other terms may also be expressed in terms of the raising and lowering operators,
\begin{eqnarray}
H_R\left(\mathbf{P},\mathbf{R}\right) &=&\alpha_R\left(\xi_+\sigma_+ a_+ - \xi_- \sigma_- a_-\right)+H.c.,~~~~~~\label{Hso-1}\\
H_D\left(\mathbf{P},\mathbf{R}\right) &=& i\alpha_D\left(\xi_+\sigma_- a_+ + \xi_- \sigma_+ a_-\right)+H.c.,~~~~~~\label{Hso-2}
\end{eqnarray}
where $\xi_{\pm}= \sqrt {m\Omega/\hbar}\pm eB/\sqrt{4m\hbar\Omega}$, $\sigma_{\pm}=\left(\sigma_x\pm i\sigma_y\right) /2$, $\omega_{\pm}=\omega\left( 1 \pm \omega_c/\left(2\omega\right)\right)$, $\Omega_{\pm}=\Omega \pm \omega_c/2$ and $\Omega=\sqrt {\omega_0^2+\omega_c^2/4}$ with $\omega_c=eB/m$  being the cyclotron frequency. In~(\ref{Hso-1}) and~(\ref{Hso-2}), $H.c.$ signifies the Hermitian conjugate. For a special case, when $B_x=B_y$, we can also write $H_{ad}$ and $H_2$ in terms of number operator which plays important role on admixture mechanism of spin states as left and right dots are pulled apart.
\begin{widetext}
\begin{eqnarray}
H_{ad}=\frac{\hbar}{2}\left(\xi_+ a_+-\xi_- a_-\right)\omega_+ x_0^{L,R}
+\frac{1}{\hbar}\left(\alpha_R -i\alpha_D \right)m\omega_+\sigma_+ x_0^{L,R} +H.c. \label{had-2}\\
H_2=\frac{e^2B_x^2}{2m_z}\left\{\ell^2\left( a_-^2 + a_-a_+^\dagger + a_+^\dagger a_- + (a_+^\dagger)^2  \right) + 2 \ell \left( a_- + a_+^\dagger\right) x_0^{L,R} + \left(x_0^2\right)^{L,R}  \right\} \label{H2-2}
\end{eqnarray}
\end{widetext}
where $x_0^L > 0$ for left dots and $x_0^R < 0$ for right dots. As can be seen in Eq.~\ref{H2-2}, the double dots Hamiltonian depends on the quadratic function of the separation between the dots. As a result in Fig.\ref{fig2}, we find parabolic bands as a function of interdot distances in a double dots. The parabolic bands crosses when double dots are separated at around $60nm$, which induces spin-hot spot, evidently seen in  Fig.\ref{fig2} (a).

In Fig.~\ref{fig3}, we plot various spin-relaxation rates versus the strength of in-plane magnetic fields in the presence of an out-of-plane magnetic field ($B_z=0.2T$) in a double quantum dot. It is remarkable to find two spin hot spots where the spin-relaxation rate shown in the green plot is several orders of magnitude smaller than that shown in the black plot. It is also interesting to find that the spin-relaxation rate shown in the green plot is insensitive to the strength of the in-plane magnetic field ($T~B^0$) until it reaches the spin hot spot. The spin-relaxation rate shown in the black plot vanishes as $B_{in}^7$. The spin-relaxation rate shown in the green plot has a spin hot spot value of around $10^3s^{-1}$, which yields a decoherence time ($T_2=2T_1$) in the millisecond range. This likely provides a suitable mechanism for spin energy transfer between the two quantum dots for qubit gate operations. The bandstructures of double quantum dots are shown in Fig.~\ref{fig3}(b). The level crossings of several bands shown in Fig.~\ref{fig3}(b) are responsible for inducing spin hot spots in Fig.~\ref{fig3}(a). The probability densities of the spin and orbital states are shown in Fig.~\ref{fig3}(c) and (d). Note that the probability densities associated with spin splitting and orbital splitting due to the applied magnetic fields are not shown in this figure, but they are identical to those shown in Fig.~\ref{fig1}(a) and (c). Finite-size effects are observed in the probability density of the orbital states, as shown in the two right columns of Fig.~\ref{fig3}(d).

In Fig.\ref{fig4}(a), we again plot various spin-relaxation rates, as in Fig.\ref{fig3}, but with the double quantum dot separation chosen to be $y_0=60$nm between their centers. As in Fig.\ref{fig3}, we again observe two spin hot spots in the vicinity of approximately $4.5T$, chosen as the amplitude of the in-plane magnetic field. Since the double dots are separated by $60$nm between their centers, we expect the formation of two distinct dots in the probability density plot. However, we observe a single dot, as shown in the bottom inset of the probability density plot, because the magnetic confinement potential is much stronger than the gate potential ($\omega_c/\omega_0$). At small magnetic fields ($B_{in} < 2T$), two additional spin hot spots are observed in the spin-relaxation rates, as shown by the black and red plots, because the band structures cross twice at around $B_{in}=1T$ and $B_{in}=1.5T$, as illustrated in Fig.\ref{fig4} (b). Such oscillations in the band structure can be attributed to the strong influence of the non-parabolicity term in the calculations, particularly when the confinement due to the gate potential is stronger than the magnetic potential. We clearly observe the formation of double quantum dots at small magnetic fields ($B_{in}=0.1T$), as shown in the upper inset of the probability density plot, because the confinement due to the gate potential is stronger than the magnetic confinement induced by the in-plane and out-of-plane magnetic fields.

\section{Conclusions}\label{conclusion}

This study shows that spin-relaxation in single and double silicon quantum dots, mediated by acoustic phonons, can be tuned using in-plane and out-of-plane magnetic fields, leading to different types of spin hot spots. In single quantum dots, the spin-relaxation rate is highly sensitive to magnetic fields at low values but converges near the spin hot spot, which arises from singlet–triplet level crossings (Fig.\ref{fig1}). In contrast, double quantum dots exhibit additional, unusual spin hot spots that emerge as the dots are spatially separated (Fig.\ref{fig2}).  Most interestingly, the decoherence time at the spin hot spot is two orders of magnitude larger than that in single quantum dots, approximately $1ns$ in Fig.\ref{fig1},  $1\mu s$ in Fig.\ref{fig2}, and $100\mu s$ in Fig.\ref{fig3}. In Fig.\ref{fig4}, it is seen that when two quantum dots are pulled apart, there is a competition between the dots formed by the gate potential and the magnetic confinement potential. As a result, two additional spin hot spots are observed at low magnetic fields ($B_{in}<2T$), and another two additional spin hot spots are observed at high magnetic fields ($B_{in}>4T$). The observed spin hot spot has a decoherence time on the order of milliseconds, which is realistic for qubit gate operations. In contrast, the other spin hot spot shown by the black and red plots exhibits an unrealistically short decoherence time on the order of picoseconds, making it unsuitable for qubit gate operations. In conclusion, we have studied several different mechanisms for inducing spin hot spots in silicon single and double quantum dots for applications in quantum information processing.

\section{Acknowledgements}\label{acknowlwdgement}

The simulations were performed at BARTIK High-Performance cluster (National Science Foundation, Grant No. CNS-1624416, USA) in Northwest Missouri State University. SP acknowledges Northwest Missouri State University, the Department of Natural Science for purchasing COMSOL multiscale multiphysics simulations software package. HSC acknowledges US National Science Foundation Grant No. PHY-2110318. RM is acknowledging the support of NSERC Discovery and CRC Programs.

\bibliography{paper25_Bib}

\end{document}